\documentclass[aps,prl,twocolumn]{revtex4}
\usepackage{graphicx}
\usepackage{amsmath}
\usepackage{amssymb}
\usepackage{braket}
\usepackage{hyperref}
\usepackage{subfigure}
\usepackage[usenames,dvipsnames]{color}

\usepackage{graphicx}
\usepackage{subfigure}
\usepackage{slashed}
\usepackage{relsize}

\newcommand{\ct}{\cite}

\newcommand{\bi}{\bibitem}
\newcommand{\be}{\begin{equation}}
\newcommand{\ee}{\end{equation}}
\newcommand{\ba}{\begin{eqnarray}}
\newcommand{\ea}{\end{eqnarray}}

\newcommand{\non}{\nonumber}

\begin{document}

\title{Mixed state dynamical quantum phase transition and emergent topology}%
\author{Utso Bhattacharya, Souvik Bandopadhyay and Amit Dutta \\
Department of Physics, Indian Institute of Technology, Kanpur-208016, India}

\title{Mixed state dynamical quantum phase transition and emergent topology}%
\author{Utso Bhattacharya, Souvik Bandopadhyay and Amit Dutta \\
Department of Physics, Indian Institute of Technology, Kanpur-208016, India}

\begin{abstract}

Preparing an integrable  system in  a mixed state described by a   thermal density matrix , we    subject it to a sudden quench and explore  the subsequent unitary dynamics. Defining a version of the generalised  Loschmidt overlap amplitude (GLOA)  through the purifications of the time evolved density matrix, we claim that non-analyiticies in the corresponding ``dynamical free energy density" persist and is referred to as mixed state dynamical quantum phase transitions (MSDQPTs). Furthermore,  these MSDQPTs are uniquely characterised by a topological index constructed by the application of the Pancharatnam geometry on the purifications of the thermal  density matrix; the quantization of this index however persists up to
a critical temperature.  These  claims are corroborated  analysing the non-equilibrium dynamics of  a transverse Ising chain initially prepared in a thermal state and subjected to
a sudden quench of the transverse field.\end{abstract}

\maketitle

Recent experimental advances in  realisation of closed condensed matter systems via cold atoms in optical lattices,  especially studies  of the real time evolution of closed quantum systems in cold atomic gases \ct{bloch08,lewenstein12,jotzu14,greiner02,kinoshita06,gring12,trotzky12,cheneau12,schreiber15} and also  light-induced non-equilibrium  superconducting and topological systems \ct{fausti11,rechtsman13} have resulted in  an upsurge in related theoretical works \ct{calabrese06,rigol08,oka09,kitagawa10,lindner11,bermudez09,patel13,thakurathi13,mukherjee09,das10,Russomanno_PRL12,bukov16,pal10,nandkishore15}. 
{(For review, we refer to \ct{dziarmaga10,polkovnikov11,dutta15,eisert15,alessio16}.)}  One of the emerging areas of research on non-equilibrium (quenched) closed quantum systems is the dynamical quantum phase transitions (DQPTs) as has been proposed by Heyl $et~al.$ \ct{heyl13}  in connection to thermal phase transitions. These phase transitions  are characterised by non-analyticities in the so called dynamical free energy during
the subsequent temporal evolution of the quenched system   when the initial state is temporally orthogonal to the time-evolved state; these non-analyticities are  manifested in the logarithm of the Loschmidt echo (or the rate function of the return probability).  Remarkably, in some situations, these non-equilibrium transitions can be characterised by a dynamical topological order parameter (DTOP) \ct{budich15} extracted   from the ``gauge-invariant" Pancharatnam phase \cite{pancharatnam56,berry84} (see [\onlinecite{zella12}] for review)
extracted from the Loschmidt overlap, i.e., the overlap between the initial state and the time evolved state.

Let us first elaborate on the basic notion of a DQPT focussing on the sudden quenching case  \ct{heyl13}.
 Denoting  the ground state of the initial Hamiltonian as  $|\psi_0\rangle$ and the final Hamiltonian reached through the
quenching process as  $H_f$, the Loschmidt overlap  amplitude (LOA) is defined as  $G(t)=\langle\psi_0|e^{-iH_ft}|\psi_0\rangle$. Generalizing $G(t)$  to 
 $G(z)$ defined in the  complex  time ($z$) plane such that $z=R+it$, 
 one can introduce the notion of a dynamical free energy density, $f(z)=-\lim_{L\to \infty} \ln{G(z)}/L^d$, where $L$ is the linear dimension of a $d$-dimensional system. 
One  then looks  for the zeros of the $G(z)$ (or non-analyticities in $f(z)$) to define a DQPT.  For a transverse Ising chain, it has been observed   that  when the system is suddenly quenched across the quantum critical point (QCP), the lines of  Fisher zeros (FZs)  cross the imaginary time axis  at  instants $t_n^{*}$; at these instants  {the rate function of the return probability} defined as $I(t) = - \ln |G(t)|^2/L$ shows sharp non-analyticities.     Several subsequent studies \ct{karrasch13,kriel14,andraschko14,canovi14,heyl14,vajna14,sharma15,heyl15,palami15,vajna15,schmitt15,budich15,sharma16,divakaran16,huang16,puskarov16,zhang16,heyl16,bhattacharya16,zunkovic16,
bhattacharya17,zvyagin17,sei17,fogarty17,halimeh171,homrighausen17}  established that similar DQPTs are observed for  sudden quenches across the QCP for both integrable and non-integrable models, although crossing the QCP is not essential in some situations \ct{vajna14,sharma15}. DQPTs have also been  observed when the final state, evolving  with the time-independent  final Hamiltonian, is prepared through a slow ramping of a  parameter of the Hamiltonian \ct{sharma16,pollmann10}. Furthermore, on the experimental side, non-analyticities have been detected in  the dynamical evolution of a fermionic many-body state after a quench \ct{flaschner16} and 
also in the non-equilibrium dynamics of  a string of ions simulating interacting transverse-field Ising models \ct{jurcevic16}.
We note in  passing that the rate function $I(t)$  is related to the Loschmidt echo which has been
studied in the context of  equilibrium quantum phase transitions and associated dynamics  \ct{quan06,rossini07,cucchietti07,venuti10,
sharma12,nag12,mukherjee12,dora13,zanardi07_echo,gambassi11,dorner12}. 

%

All these afore mentioned studies concern the subsequent  unitary dynamics of the initially prepared pure state $|\psi_0\rangle$ following a quench of the system. The questions we address are the following: does the fascinating aspects of DQPTs arising
due to quantum coherence get wiped  out if the initial state happens to be a thermal (mixed) state. To achieve this goal, we prepare the initial state in a thermal Gibbs state described by
a density matrix  $\rho_0$
characterised by an inverse  temperature $\beta$; a parameter of the Hamiltonian is then suddenly changed and we propose
a generic  form of the LOA (that reduces to the conventional definition given above in the pure state limit) to track the subsequent unitary evolution. We then explore the following questions: do mixed state  DQPTs (MSDQPTs), manifested in the zeros of  the generalized LOA (GLOA),  occur in this situation? If so, can these  MSDQPTs be characterised by an appropriate topological index? Interestingly, we find affirmative answers for all these questions.  MSDQPTs indeed
exist. Furthermore, these non-equilibrium transitions are characterized by a gauge-invariant topological index up to a sufficiently high initial  temperature (denoted by $\beta_c$)  below which the quantisation gets destroyed.  In passing, we note that there have been several studies in recent years which explore the interconnection between temperature and topology
of  equilibrium systems \cite{delgado13,arovas14} using Uhlmann phase \ct{uhlmann86}.
However, the topological index that we use here to characterise the MSDQPTs manifests out of  the  interferometric phase \ct{sjoqvist00}. 

\noindent {\it Quenches, GLOA and MSDQPTs:} Let us consider a many-body integrable  Hamiltonian that can be decoupled   into  $2\times2$ Hamiltonians  for each  momentum mode $k$, i,e.
$\hat{H}_k=\vec{d}_k \cdot {\vec {\sigma}}$, where  ${\sigma}_i$s are the Pauli matrices.  Henceforth,  we set $\hbar$ and the
Boltzmann constant  $k_B$ to unity. 
The mixed state density matrix at time $t=0$ describing the system  at  thermal equilibrium with a bath corresponding to the initial Hamiltonian 
$\hat{H}^i_{k}=\vec{d}^i_{k}\cdot {\vec{\sigma}}$ can be written as

\begin{equation}
{\hat \rho}_k(0)=\frac{e^{-\beta {\hat H}^i_{k}}}{{\rm Tr}(e^{-\beta {\hat H}^i_{k}})}=\frac{1}{2}(\mathbb{I}_k-n{\hat{{d}}^i}_{k}\cdot {\vec {\sigma}})
\end{equation}
where $\beta$ is the inverse temperature,
$ n=\tanh(\beta \epsilon^i_k)$, $\hat{{d}}^i_{k}= \vec{d}^i_k/\epsilon^i_k$ with   $\epsilon^i_k=|\vec{d^i_{k}}|$ and $\mathbb{I}_k$ is the $2\times 2$ identity matrix.

 Having prepared the initial density matrix we decouple the system from the bath \ct{dorner12} at $t=0$ and subject it  to a sudden change in a parameter $h$ of the Hamiltonian   from an initial value $h_i$ to a final value $h_f$; we then probe the subsequent non-equilibrium dynamics of the system
 generated by the final time-independent Hamiltonian
$ \hat{H}^f_{k}=\vec{d}^f_{k}\cdot \hat{\vec{\sigma}}$,
 so that the density matrix ${\hat \rho}_k(t)$ at a time $t$ after the quench is given by ${\hat \rho}_k(t)={\hat U}_k(t) \hat {\rho}_k(0){\hat U}_k^{\dagger}(t)$.

\begin{figure}
\begin{center}
\includegraphics[width=.20\textwidth,height=3cm]{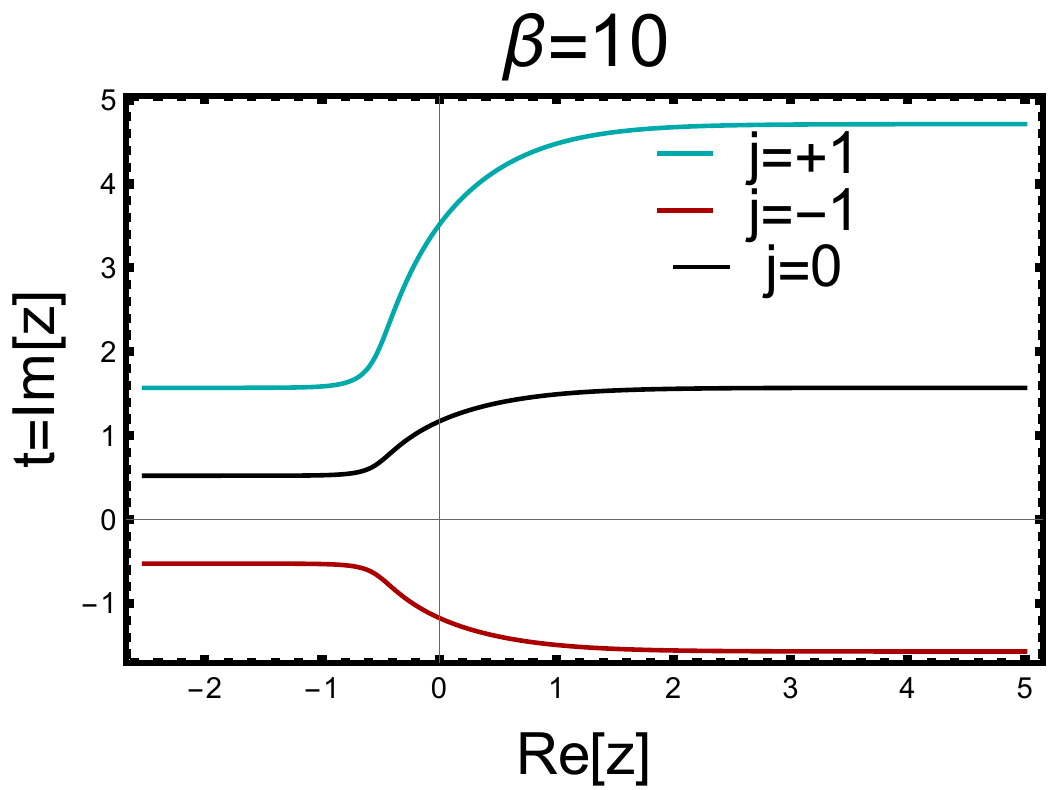}
\includegraphics[width=.20\textwidth,height=3cm]{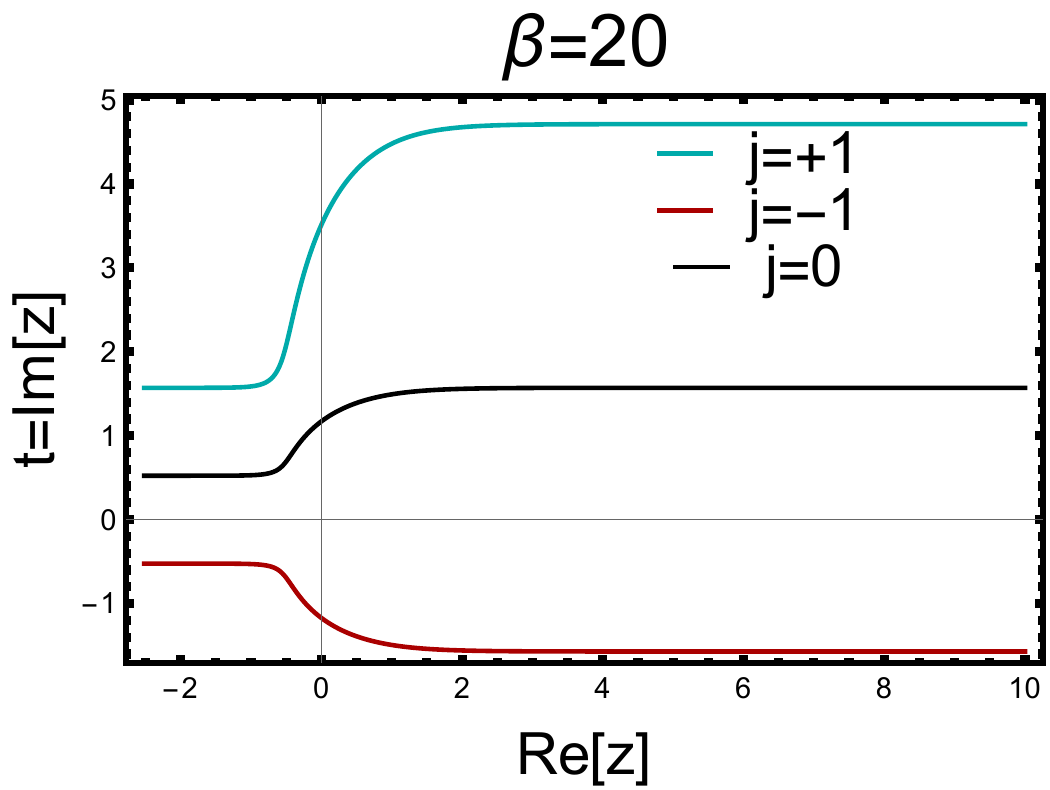}
\caption{ The lines of FZs in the complex $z$ plane following a quench from the initial value of the transverse  $h_i=0.5$ and to the final value $h_f=2$ across the QCP $h=1$ for the Hamiltonian
\eqref{eq_ham}  corresponding
to the initial $\beta=10$ and $\beta =20$. The lines of FZs cut the imaginary (real time) axis at critical times $t_j^*$ given in Eq.~\eqref{eq_time} corresponding to $j=0,\pm 1$.}
\label{fig_FZs}
\end{center}
\end{figure}

 Since we are dealing with the dynamics of mixed state, one needs to define a generalised Loschmidt overlap amplitude (GLOA) defined  for each $k$ mode as the overlap between the purifications \ct{uhlmann86} as
 (for a discussion on the purifications, refer to the supplementary materials (SM))
 \begin{eqnarray}
 L_k&=&\langle w_k(0)|w_k(t)\rangle=Tr ({\hat {\rho}}_k(0) {\hat U}_k(t)) ~~\\
 &{\rm where}&~~{\hat U}_k(t)=e^{-i {\hat H}^f_{k}t}~~~{\rm and}~~{\cal  L} =\prod L_k,
  \end{eqnarray}
 where ${\cal L}$ is called the total GLOA obtained through a product  over all the modes $k$ from $0$ to $\pi$. Notably, the GLOA defined for the mixed state situation reduces to the conventional form of the LOA,  $\langle \psi_0|e^{-i{\hat H}_f t}|\psi_0\rangle$,
 in the pure state limit when the system is in the initial state $|\psi_0\rangle$, before the quenching is applied. 
 
The complete gauge invariant phase obtained from the GLOA for each $k$ mode is given by \ct{sjoqvist00}, 
\begin{eqnarray}
\phi^k_{g}&=&\phi^k-\phi^k_{dyn}\non \\
&=&{\rm Arg} [{\rm Tr}(\hat{\rho_k}(0) {{\hat U}_k}(t))]+\int_0^t {\rm Tr}(\hat{\rho_k}(t)\hat{H}^f_k)dt
\label{eq_geo_phase}
\end{eqnarray}
where $\phi^k={\rm Arg} [{\rm Tr}
(\hat{\rho_k}(0)\hat{U_k}(t))]$ and $\phi^k_{dyn}= -\int_0^t {\rm Tr}(\hat{\rho_k}(t)\hat{H}^f_k)dt$. See the SM for a thorough discussion.
 For the sudden quench protocol   from an initial value of the parameter  $h_i$ to a  final value $h_f$ with initial state characterised by the initial inverse  temperature $\beta$, we find

\begin{equation}
\phi^k=\tan ^{-1}\left[\frac{n\vec{d}^i_{k} \cdot \vec{d}^f_{k}}{\epsilon^i_k.\epsilon_f^k}\tan (\epsilon_f^kt)\right];~~~
\phi^k_{dyn}=\frac{n\vec{d}^i_{k} \cdot \vec{d}^f_{k}t}{\epsilon^i_k},
\label{eq_phases}
\end{equation}
where $\epsilon_{k}^{f}=|\vec{d}^f_{k^{}}|$.
\begin{figure*}
\begin{center}
\includegraphics[width=0.37\textwidth]{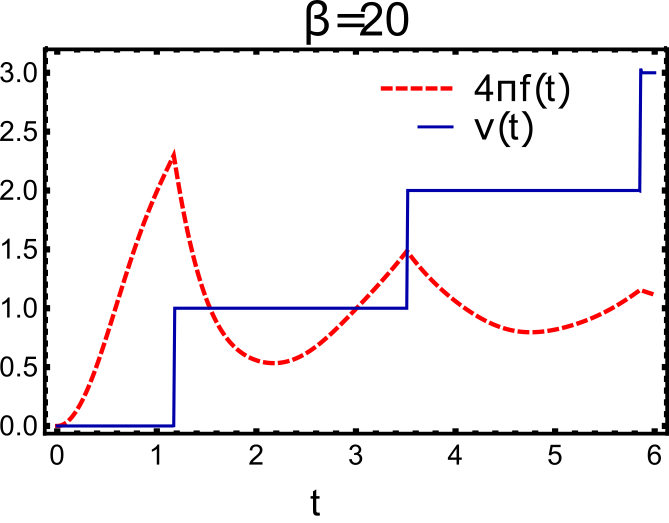}
\includegraphics[width=0.37\textwidth]{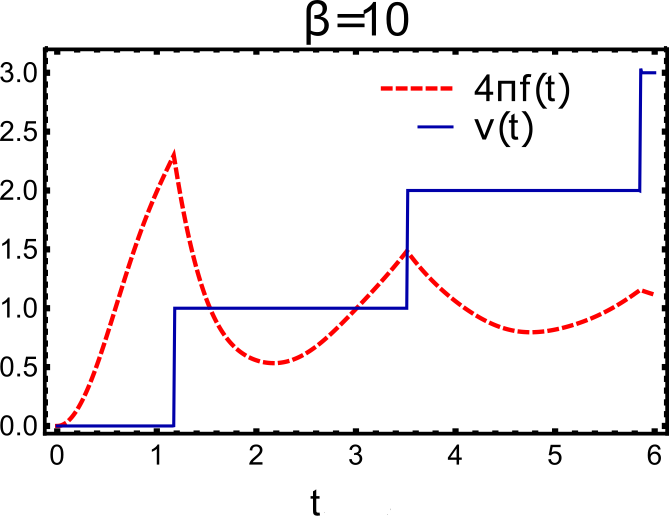}
\caption[Fig.1]{The figures show non-analyticities in the dynamical free energy and the jump in the  MSDTOP $\nu (t)$ as functions of time  at the instants of real time shown in Fig.~\ref{fig_FZs} after a sudden quench from $h_i=0.5$  to $h_f=2.0$ in Hamilonian \eqref{eq_ham}. We have scaled $f(t)$ by a factor of $4\pi$ to superimpose its temporal variation on that of  $\nu (t)$. }
\label{fig_dtop}
\end{center}
\end{figure*}
  
 Let us now proceed  to  define the dynamical counterpart of the free energy density in the thermodynamic limit analogous to the pure state situation \ct{heyl13},
  
   \begin{equation}
 f(t)=-\frac{1}{2\pi}{\rm Re} \left[\int_{0}^{\pi}\log[{\rm Tr}(\hat{\rho}_k(0) {\hat U}_k(t))]dk\right].
 \label{eq_free_energy}
\end{equation}\\

\noindent  The position of the non-analyticities in the free energy density can be determined solely by the zeros of the partition function $L_{k}(z)=Tr[\hat{\rho}_{k}(0)e^{-\hat{H}^f_{k} z}]$
(where $z\in C$); these are reflected in the real time evolution  of the dynamical free energy density when the zeros  lie on the imaginary (real time axis).
The zeros of $L_k(z)$ can be evaluated for the present case as
\begin{equation}
z_j(k)=\frac{1}{2\epsilon^f_{k}}\ln\left(\frac{1-n\hat{d}^i_{k}\cdot \hat{d}^f_{k}}{1+n\hat{d}^i_{k}\cdot \hat{d}^f_{k}}\right)+\frac{i(2j+1)\pi}{2\epsilon^f_{k}}
\end{equation}
where   ${\hat d}^f_{k} = \vec{d}^f_{k}/\epsilon^f_{k}$  and $j \in \mathbb{Z}$.
The line of FZs in the complex plane cuts the imaginary axis for the mode $k^{*}$ when $\hat{d}^i_{k^{*}}\cdot \hat{d}^f_{k^{*}}=0$; consequently, one
observes  non-analyticities in the dynamical free energy density  $f(t)$ at   instants of real time given by:

\begin{equation}
t_{j}^{*}=\frac{(2j+1)\pi}{2\epsilon_{k*}^{f}}
\label{eq_time}
\end{equation}
where $\epsilon_{k*}^{f}=|\vec{d}^f_{k^{*}}|$. Evidently, the critical times given in Eq.~\eqref{eq_time} do not depend upon the initial temperature of the system.

\noindent {\it Mixed state dynamical topological order parameter (MSDTOP):} 
Let us now address the question whether  the MSDQPTs  occurring at critical times $t_j^*$  can be characterised by an appropriate 
topological index.   To address this question, in a spirit similar to the pure state situation \ct{budich15},  we invoke upon  the gauge invariant geometric phase defined in \eqref{eq_geo_phase} to construct the MSDTOP as
\begin{equation}
\nu (t)=\frac{1}{2\pi}\int_{0}^{\pi}\frac{\partial \phi_g^k}{\partial k}dk
\label{eq_dtop}
\end{equation}
The MSDTOP will remain fixed between two critical times  showing   discontinuous jumps at every MSDQPT thereby appropriately characterising them
as long as the initial temperature does not exceed the minimum characteristic energy scale of the problem. 
In the pure state limit evidently  $\phi^k_g$ reduces to the Pancharatnam geometric phase and one can readily check that for the modes
 $k=0$ and $k=\pi$, $\phi^{k}_g$ is pinned to zero \ct{budich15}. Even when  $\beta$ is finite but $\epsilon^i_k \gg T$ (so that $n \to 1$),
 $\phi^k_g$ still remains periodic in the interval $k\in[0,\pi]$; consequently, $\nu(t)$ cannot change  in the region $k\in[0,\pi]$  except for the instants 
 of MSDQPTs  when, as seen from Eq.~\eqref{eq_phases},  the phase $\phi^k$ becomes ill-defined. 
We  therefore find that  MSDTOP $\nu(t)$  stays quantised and jumps by a factor of unity at every MSDQPT. However if  $\epsilon^i_k\lesssim T$, so that $n \lesssim 1$ the phase  $\phi^k_g$ is no longer periodic in the interval $k\in[0,\pi]$, rather we have
 \begin{eqnarray}
 \nonumber\phi_g^{k=0,\pi}&=&\tan ^{-1}\left[n(k=0,\pi)\tan (\epsilon^f_{k=0,\pi}t)\right]-\\
 &&n(k=0,\pi)\epsilon^f_{k=0,\pi}t
 \end{eqnarray}
 where the quantity $n$ also depends on $k$ rendering  $\phi^{k=0}_g \neq \phi^{k=\pi}_g$ and hence destroying  the periodicity in $\phi^{k}_g$.
 Thus, we can define a critical initial temperature $T_c$ above which the quantization of the MSDTOP gets destroyed, determined by the condition,
 \begin{equation}
 \tanh\left(\frac{\epsilon_i^{k^{'}}}{T_c}\right)\sim1
 \label{eq_critical}
 \end{equation}
 where $k^{'}$ is the mode for which  the minimum gap in the spectrum of  $\epsilon^i_k$  is $\epsilon^i_{k'}$.

\begin{figure}
\begin{center}
\includegraphics[width=0.37\textwidth]{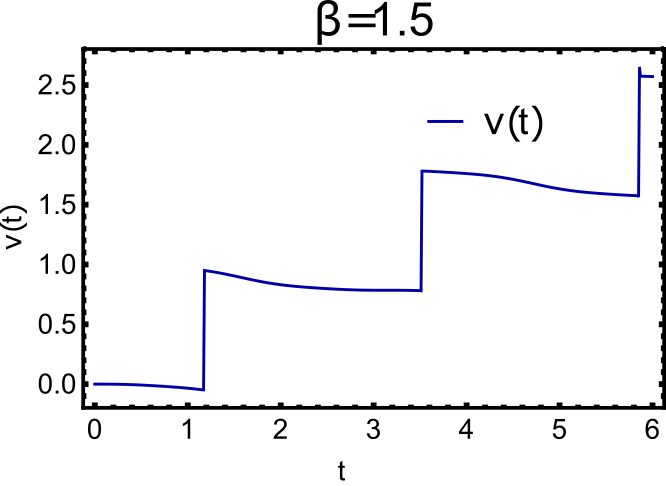}
\caption{The perfect quantisation of $\nu(t)$ between two successive critical times for the same quenching protocol as shown in Fig.~\ref{fig_dtop}   gets destroyed when the initial temperature is relatively higher. }
\label{fig_finite_T}
\end{center}
\end{figure}

\noindent {\it Illustration with the  transverse field Ising chain: }
We shall now illustrate the above claims considering the integrable one-dimensional (1D) transverse Ising Hamiltonian \cite{sachdev96, sei13, dutta15}:

\begin{equation}
\hat{H}=-J\sum_{\langle i,j\rangle} {\sigma}^x_i {\sigma}^x_j-h\sum_i {\sigma}^z_i
\label{eq_ham}
\end{equation}
where  $J$ (henceforth set equal to $1$)  is the nearest neighbour ferromagnetic coupling strength and $h$ is the  non-commuting transverse magnetic field.  Employing a   Fourier transformation and  a subsequent    Jordan-Wigner transformation, 
one can map  the many-body Hamiltonian    into  decoupled  $2\times2$  Hamiltonians  for each  momentum mode $k$, i,e.
$\hat{H}_k=\vec{d}_k \cdot {\vec{\sigma}}$, where the ${\vec d}_k$ has components  $(0,\cos k-h,\sin k)$. Analysing the spectrum of the model, one can show 
that the TFIM has two Quantum Critical Points (QCPs) at $h = \pm 1$ separating the ferromagnetic phase from  the paramagnetic phases for $|h| >1$. \\
Having prepared the system in a thermal state, we employ a sudden change in the field $h$ from $h_i$ to $h_f$ and hence in the present example,
 $\vec{d}^i_{k}\equiv(0,\cos k-h_i,\sin k)$ and $\vec{d}^f_{k}\equiv(0,\cos k-h_f,\sin k)$.  In the following, we choose   $h_i=0.5$ and  $h_f=2$ so that the spin chain is quenched across
 the QCP at $h=1$. We indeed find that the lines of FZs cross the real time axis at critical times  $t_{j}^{*}$ as shown in  Fig.~\ref{fig_FZs} where  the dynamical free energy density  $f(t)$ defined in Eq. \eqref{eq_free_energy} becomes non-analytic (as shown in Fig.~\ref{fig_dtop}). Moreover, here we also show that the MSDTOP defined in Eq.~\eqref{eq_dtop}  remains  fixed between two critical times  showing  discontinuous jumps by a factor
 of unity at every MSDQPT, thereby appropriately characterising them. However, if the initial temperature is relatively high, as we illustrate in Fig.~\ref{fig_finite_T}, the perfect quantisation of MSDTOP gets destroyed.
 The critical condition follows from Eq. ~\eqref{eq_critical} with $k^{'}=0$ in the present example.

\medskip
\noindent {\it Experimental possibilities:} The first experimental observation of quantum geometric phases for mixed states was performed by Du $et~al.$ \ct{du03}. The quantum holonomies for a mixed state of spin half nucleus was measured and observed in accordance with the theoretical predictions by Sjoqvist $et~al$ \ct{sjoqvist00}. Interestingly, for a pure state non-equilibrium dynamics, DQPTs were observed using state tomography technique, where Flaschner $et~al.$ \ct{flaschner16} measured the time evolution of the pure state for each quasi-momenta mode on the Bloch sphere to reconstruct the instantaneous many-body state. Combining this technique with the experimental procedure of Ref.~[\onlinecite{du03}], a similar approach may be undertaken to map a time evolving mixed state on a Bloch sphere to measure the geometric phase acquired and observe MSDQPTs. Furthermore, the successful measurement of DQPTs in an interacting many-body system such as the transverse field Ising model with long range interactions, opens up the possibility of the observation of MSDQPTs and MSDTOPs in such a model.
  
\noindent {\it Concluding comments:} We have defined a GLOA for a thermal mixed state undergoing unitary dynamics using purifications and have proposed a MSDTOP similar to the construction of the gauge invariant interferometric phase. Our GLOA and MSDTOP comes out of a generic framework which is independent of the underlying integrable model involved.  To substantiate our claims we take the example of the 1-D Ising model and show the existence of MSDQPTs and a MSDTOP.
 We observe that although the thermal nature of the mixed state completely destroys the quantization of the MSDTOP when the initial temperature scale becomes comparable to the temperature associated with the initial minimum energy gap, the critical modes and times of the MSDQPTs which depend only on the unitarity of the quenching process involved, remain unaffected. This firmly establishes the existence of a critical temperature below which MSDQPTs exist and are characterized by the MSDTOP just as in the $T\rightarrow0$ limit. Since, our formalism holds true for integrable  quantum systems, the existence or non-existence of MSDQPTS for any general unitary dynamics involved can be confirmed. It also has a natural well defined reduction to the pure state limit. Although the definition of GLOA is indeed independent
 of the dimensionality of the system, the MSDTOP defined here is strictly valid for a one dimensional system though may even be generalised to higher dimensions
 \ct{bhattacharya17}.

\widetext
\begin{center}
\textbf{\large  ``Supplementary Material on ``Mixed state dynamical quantum phase transition and emergent topology''''}\\
\vspace{0.5cm}
{ Utso Bhattacharya, Souvik Bandopadhyay and Amit Dutta  }\\
\vspace{0.2cm}
{}{\it Indian Institute of Technology Kanpur, Kanpur 208 016, India} \\
\end{center}

\setcounter{equation}{0}
\setcounter{figure}{0}
\setcounter{table}{0}
\setcounter{page}{1}
\makeatletter
\renewcommand{\theequation}{S\arabic{equation}}
\renewcommand{\thefigure}{S\arabic{figure}}
\renewcommand{\bibnumfmt}[1]{[S#1]}
\renewcommand{\@cite}[1]{[S#1]}

 The mathematical extension of the Pancharatnam-Berry phase \cite{pancharatnam56}, \cite{berry84}  and the whole underlying geometry of pure
quantum states (see \onlinecite{zella12} for review) can be extended to mixed states by a redefined parallel transport criteria for mixed state density matrices, defined in the extended Hilbert Space of the purifications of the density matrix \cite{uhlmann86}, \cite{uhlmann89}. In general there are two approaches i) gauge invariant phase for density matrices defined by Uhlmann \cite{uhlmann86}  which has been used to study finite temperature topologies for thermal mixed states  at equilibrium \cite{viyuela14}. (ii) The gauge invariant interferometric phase defined more recently based on purifications of density matrices
\cite{sjoqvist00}.

\section   {Purifications:}
Purification of a density matrix is defined on an extended Hilbert space constructed by taking the direct product of the original Hilbert space  ${\cal H}_0$  with an {\it ancillary} Hilbert space ${\cal H}_A$ (whose states do not transform under the operators defined on ${\cal H}_0$). Let there be a density operator $ {\rho}$ defined to act on the Hilbert space ${\cal H}_0$. It can be expanded in it's eigen-basis as: $  {\rho}=\sum_{i} p_i|\psi_i\rangle\langle\psi_i|$.
 Now, a normalized state (purification)  $|w\rangle\in {\cal H}_p (={\cal H}_{0} \otimes {\cal H}_A)$ of $ {\rho}$ is defined as \ct{uhlmann86}
 
 \begin{equation}
|w\rangle=\sum_{i}\sqrt{p_i}| \psi_i\rangle\otimes |\psi_i' \rangle, ~~~{\rm where}~~~ |\psi_i'\rangle\in H_A,
\end{equation} 
so that the original density matrix can be obtained by tracing out the ancillary states
$ {\rho}=Tr_A(|w\rangle\langle|w|)$.
Under a unitary transformation $U$,  the state $|\psi_i(t)\rangle\rightarrow U_0(t)|\psi_i(0)\rangle$.  On the contrary,
the purifications transform as
\begin{equation}
|w(t)\rangle=\sum_{i}\sqrt{p_i}| \psi_i(t)\rangle\otimes |\psi_i' \rangle=U(t)\sum_{i}\sqrt{p_i}| \psi_i(0)\rangle\otimes |\psi_i' \rangle
\label{eq_purification}
\end{equation}
where $U(t)=U_0(t)\otimes I_A$ and $I_A$ is the identity operator of $H_A$. Comparing with the main text, the index $i$ refers to the two states of the density matrix corrresponding to each mode $k$.

\section{The  topological structure:}
Imposing the constraint that  $|w(t)\rangle$ are normalized leads to the condition $\frac{d}{dt}\langle w(t)|w(t)\rangle=0$.
Now the complete formulation of Pancharatnam and Berry\cite{pancharatnam56,berry84} can be extended to these purifications   as was shown by Sj\"{o}qvist et al \ct{sjoqvist00} which leads to a topological index different from that introduced by Uhlmann \ct{uhlmann86,uhlmann89}.  Let the states $|w(t)\rangle$ be parametrized by a continuous parameter $t$, with $|w(t)\rangle$ describing a curve in the Hilbert space ${\cal H}_p$.
A metric is defined in $H_p$ as the measure of distance between two states as  
$d=|| |w(t_1)\rangle-|w(t_2)\rangle||$.

Let us note that  two states are said to be parallel if the distance between them is minimum. But, the purification states $|w(t)\rangle$ also have a phase ambiguity or an $U(1)$ gauge freedom as a gauge transformation $ |w(t)\rangle\rightarrow e^{i\phi}|w(t)\rangle$
produces the same density matrix and preserves inner products in the space ${\cal H}_p$.
 Now, this phase ambiguity or a $U(1)$ gauge choice needs to be fixed in order to define unique trajectories of $|w(t)\rangle$ in ${\cal H}_p= {\cal H}_0 \bigotimes  {\cal H}_A$. In the  parallel transport criteria imposed by  Pancharatnam\cite{pancharatnam56,berry84} for pure states,  one  fixes the gauge at every point such that two infinitesimally separated states are parallel to each other.
 For the purifications,  the corresponding  parallel transport condition can be recast  to the form \ct{sjoqvist00}:
 \begin{equation}
 \langle w(t)|\dot{w}(t)\rangle=0, \implies  Tr( {\rho}(0)U^{\dagger}\dot{U})=0,
 \end{equation}  
where in the second step  we have used Eq.~\eqref{eq_purification} and 
$ {\rho}(0)=\sum_{i} p_i | \psi_i (0) \rangle\langle \psi_i (0)|$.

The parallel transport condition fixes the phase at every point of the trajectory in ${\cal H}_p $ and for a  transport in time  from $0$ to $t$  is given by

\begin{equation}
\phi={\rm Arg} \langle w(t_2)|w(t_1)\rangle \implies  \phi={\rm Arg} \left[ {\rm Tr}\left({  \rho}(0) {U}(t)\right) \right],
\end{equation}
which is the associated Pancharatnam phase. We would like to emphasize,  that this phase becomes gauge invariant only when the unitary operator   $ {U(t)}$ ensures a parallel transport. 
Now, consider an {\it open} but  unitary trajectory being generated by $U(t)$ which can be viewed as  the unitary time evolution generated by 
 $H$ (which in the main text becomes the time-independent final Hamiltonian). As in the pure case \ct{berry84,zella12}, the unitary operation
 $U$  introduces a {\it dynamical phase} to the purifications in addition to the geometric phase;  this dynamical phase  must be subtracted from the total phase rendering the geometric part   gauge invariant. 
The dynamical phase accumulated by the purifications over a time $t$ can easily be written as:
\begin{equation}
\phi_{\rm dyn}=-\int_0^t \langle w(t)| {H}|w(t)\rangle dt = -\int_0^t {\rm Tr}( {\rho(t) } {H})dt
\end{equation}
This amounts to  a weighted average of dynamical phases of the pure eigenstates of the density matrix and hence, evidently correctly reproduces
the pure state limit.  The gauge invariant (purely)  geometrical phase following  a unitary time evolution of a density matrix 
\begin{equation}
\phi_{g}=\phi-\phi_{\rm dyn}={\rm Arg} \left[ {\rm Tr}\left({  \rho}(0) {U}(t)\right) \right]+\int_0^t {\rm Tr}( {\rho(t)} {H})dt
\label{s_eq_phiG}
\end{equation}
The conclusion that this phase is gauge invariant follows from the fact that it is induced by a manifestly gauge invariant phase, namely, the Pancharatnam-Berry phase for the purifications $|w(t)\rangle$ and reduces to the pure geometric phase if the parallel transport condition is met (as in this case, the second term in Eq.~\eqref{s_eq_phiG} vanishes). In the pure state limit, the above phase reduces to the Pancharatnam-Berry phase
\begin{equation}
\phi_{\rm pure}={\rm Arg}\langle\psi(0)|\psi(t)\rangle+\int_{0}^{t}\langle\psi(t)| {H}|\psi(t)\rangle dt
\end{equation}
where the pure state density operator, $ {\rho}(t)=|\psi(t)\rangle\langle\psi(t)|$. In the main text, we have considered the temporal evolution
of the initial density matrix (characterising a mixed state) generated by the final time-independent Hamiltonian and calculated $\phi$ and
$\phi_{\rm dyn}$ for each momentum mode $k$.


\vspace{-\baselineskip}

\end{document}